# Noise in the wire: the real impact of wire resistance for the Johnson (-like) noise based secure communicator

Laszlo B. Kish [a] and Jacob Scheuer [b,c]

[a] *Department of Electrical and Computer Engineering, Texas A&M University, College Station, TX 77843-3128, USA*

[b] *School of Electrical Engineering, Tel-Aviv University, Ramat-Aviv, Israel*

[c] *Center for Nanoscience and Nanotechnology, Tel-Aviv University, Ramat-Aviv, Israel*

(second version March 9, 2010)

**Abstract.** We re-evaluate the impact of wire resistance on the noise voltage and current in the Johnson-(like)-noise based secure communicator, correcting the result presented in [*Physics Letters A* **359** (2006) 737]. The analysis shown here is based on the fluctuation-dissipation and the linear response theorems. The results indicate that the impact of wire resistance in practical communicators is significantly lower than the previous estimation.

## 1. Introduction

Following the introduction of the Kirchhoff Law and Johnson (-like) Noise (KLJN) cipher [1], it was suggested that the system was vulnerable to time domain analysis of the currents and voltages evolution in the system [2, 3] as well as to the voltage drop on the non-ideal wire [2, 4]. The time domain claims and related statistical issues were answered in [5,6]. In the present paper, we revisit the analysis of the impact of the wire resistance, showing that in fact the expression derived in [2] was inaccurate. Here we present the correct derivation of the problem of wire resistance.

Before the analysis we would like to emphasize the following remarks:

a) In the experimental realization of the KJLN communicator by Mingesz, et al [7], the wire resistance was 2% of the sum of Alice's Bob's driving resistances. Based on the measured voltage distributions, the raw information leak accessible by Eve in this case (before privacy amplification) was estimated to be 0.19% of the key bits, a value which is substantially better than in any typical quantum key distribution. This value indicates a much higher security than practical quantum mechanical based systems where Eve has utilized proper non-idealities for attack [8-10].

b) We emphasize that, the fully protected version of the KLJN key exchanger [7,11], is less susceptible to any attack based on detecting differences in the voltage or current along the wire. Alice and Bob are continuously measuring the voltage and current data at the two ends of the line, and are broadcasting and comparing these data via public channels [7, 11]. Thus they exactly know all the information accessible to Eve and can act accordingly [6] by discarding key bits where security has been compromised beyond a certain threshold. Just for comparison, QKD systems also offer some relevant protection but at a lower efficiency level because it necessitates the evaluation of the *error statistics* over a sequence of bits, an operation that requires a long record. However, in the KLJN protocol, Alice and Bob have the full knowledge of Eve's best guess *for each bit* [6], thus knowing if this guess is correct or not. This is in contrast to QKD (where Alice and Bob do not know Eve's measurement outcome), and originating from the classical physical nature and robustness of the information transfer [6].

## 2. The results in [2] and their implications

According to Eqs. (12) in [2] the voltage and current power density spectra measured by Eve at an asymmetric point in the wire [2] with wire resistances $R_{W1}$ and $R_{W2}$ toward Alice and Bob, respectively ($R_{W1} + R_{W2} = R_W$) are:

$$S_{ch}(f) = \frac{4kT[R_A(R_B + R_{W2}) + R_B(R_A + R_{W1})]}{R_A + R_B + R_{W2} + R_{W1}} \qquad (1)$$

$$S_i(f) = \frac{4kT(R_A + R_B)}{R_A + R_B + R_{W2} + R_{W1}}, \qquad (2)$$

where $S_{ch}(f)$ is the power density spectrum of the channel voltage at the given point in the wire, $S_i(f)$ is the power density spectrum of the current in the wire (loop current), $R_A$ is Alice's resistance and $R_B$ is Bob's resistance, $k$ is the Boltzmann constant, and $T$ is the effective absolute temperature of the noise generators of Alice and Bob. The Johnson noise of the wire is neglected because $T$ is orders of magnitudes greater than the ambient temperature, corresponding to an external voltage noise generator.

To observe the highest voltage drop on the wire, that is, to extract the greatest amount of information, Eve must measure at the two ends of the wire and compare the voltages.

Thus, for the sake of simplicity, we suppose: $R_{W1} = 0$ and $R_{W2} = R_W$. Eq. (1) then yields that the noise voltage at Alice's end is:

$$S_{ch,A}(f) = \frac{4kT[R_A(R_B + R_W) + R_B R_A]}{R_A + R_B + R_W} \qquad (3)$$

Applying Eq. 3 for Bob's end of the channel we use $R_{W1} = R_W$ and $R_{W2} = 0$ thus:

$$S_{ch,B}(f) = \frac{4kT[R_A R_B + R_B(R_A + R_W)]}{R_A + R_B + R_W}, \qquad (4)$$

Finally, Eve extracts the most information when she measures at both ends and compares the voltages. Note, that in the practical run, Eve does not even have to measure anything because the current and voltage values measured at the two ends are broadcasted (and compared) via public channels by Alice and Bob for a full protection against invasive attacks [6,11]. This also means, as we have already mentioned, that Alice and Bob are well aware of the actual voltage drop and the best guess of the bits by Eve [6].

Eve can extract significant information only when (the voltage-based alarm [11] is off and) the difference between the mean-square of the noise voltages (spectra integrated over the whole frequency range of operation) at the two ends is approaching or greater than the statistical inaccuracies of them during the finite sampling which is limited by the duration of the clock period. The difference between the spectra is:

$$S_{u,A-B}(f) = \frac{4kT[R_W(R_A - R_B)]}{R_A + R_B + R_W} \tag{5}$$

## 3. Analysis of the results of [2] shown above

Referring to Eqs. (3, 4), we note that in the case of zero wire resistance $R_W = 0$, the voltage noise spectrum is twice as large than what can be expected of a system comprising a set of two resistors $R_A$ and $R_B$ connected in parallel. This fact indicates that the results in [2] are inaccurate. The correct result, shown below, indicates significantly less information for Eve than the relevant result in [2], as shown in Section 5.

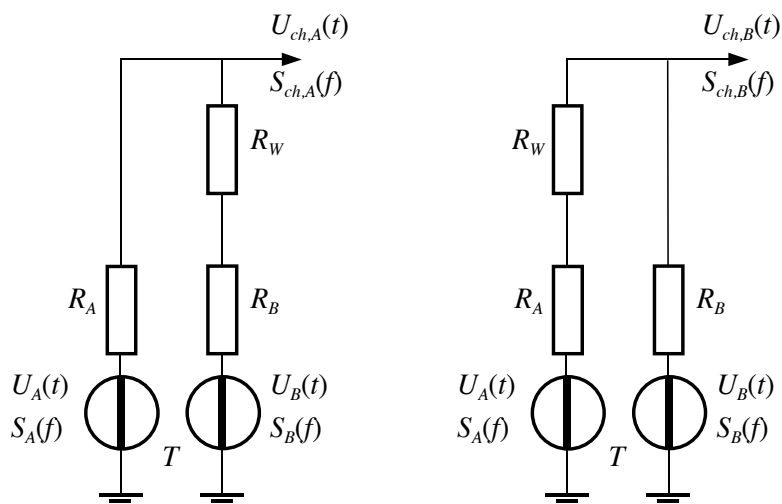

**Figure 1.** Circuit model to calculate the power density spectra of the channel voltage at Alice's and Bob's ends (left and right circuits). Similarly to the assumptions of [2], the effective temperature $T$ of the voltage generators is supposed to be so high that the thermal noise of the wire is negligible.

## 4. The wire resistance problem

In Figure 1, the model circuits for the straightforward calculation of the voltage noise spectra at Alice's end (left panel) and Bob's ends (right panel) are shown. Alice's voltage generator has $U_A(t)$ voltage amplitude and $S_A(f) = 4kTR_A$ spectrum, where $T$ is the effective temperature. Similarly, Bob's voltage generator has $U_B(t)$ voltage amplitude and $S_B(f) = 4kTR_B$ spectrum. Similarly to [2], the effective temperature $T$ of the voltage generators is assumed to be substantially higher than the thermal noise of the wire, which is, therefore, neglected.

The instantaneous channel voltage amplitude at Alice's end of the wire is the linear superposition of Alice's and Bob's generator voltages after reduced by appropriate voltage divider terms:

$$U_{ch,A}(t) = U_A(t) \frac{R_W + R_B}{R_A + R_W + R_B} + U_B(t) \frac{R_A}{R_A + R_W + R_B} \tag{6}$$

Similarly, the channel voltage at Bob's end:

$$U_{ch,B}(t) = U_B(t) \frac{R_W + R_A}{R_A + R_W + R_B} + U_A(t) \frac{R_B}{R_A + R_W + R_B} \tag{7}$$

Because the two noise voltages are statistically independent, the rule of linear operations on noises can be applied to get the separate noise spectra, and no cross-term appears:

$$S_{ch,A}(f) = S_A(f) \left( \frac{R_W + R_B}{R_A + R_W + R_B} \right)^2 + S_B(f) \left( \frac{R_A}{R_A + R_W + R_B} \right)^2 =$$

$$= 4kT_{eff} \left[ R_A \left( \frac{R_W + R_B}{R_A + R_W + R_B} \right)^2 + R_B \left( \frac{R_A}{R_A + R_W + R_B} \right)^2 \right] \tag{8}$$

and

$$S_{ch,B}(f) = S_B(f) \left( \frac{R_W + R_A}{R_A + R_W + R_B} \right)^2 + S_A(f) \left( \frac{R_B}{R_A + R_W + R_B} \right)^2 =$$

$$= 4kT_{eff} \left[ R_B \left( \frac{R_W + R_A}{R_A + R_W + R_B} \right)^2 + R_A \left( \frac{R_B}{R_A + R_W + R_B} \right)^2 \right] \tag{9}$$

where $T_{eff}$ is the effective temperature of the noise sources ($T_{eff} >> T$). We note that for the zero wire resistance case Eqs. (8, 9) do provide the correct expression for the Johnson noise: $S_{ch}(f) = 4kT_{eff} R_A R_B / (R_A + R_B)$.

The difference between the measured voltage noise spectra at the two ends is the difference between (8) and (9):

$$S_{ch,A-B}(f) = 4kT_{eff}\left[\frac{R_W^2(R_A - R_B)}{(R_A + R_W + R_B)^2}\right] \quad (10)$$

## 5. Comparison of the results of [2] and that of the current derivation

Let us suppose, that at a given secure bit exchange, Alice uses the larger resistor and Bob uses the smaller one. For simplicity, we assume the following relations between the resistances:

$$R_W \ll R_B \ll R_A \quad (11)$$

Then, for the difference of the spectra at the two ends, according to Eq. (5), yields:

$$S^{[2]}_{ch,A-B}(f) \approx 4kT_{eff}R_W \quad , \quad (12)$$

while (10) yields:

$$S_{ch,A-B}(f) \approx 4kT_{eff}\frac{R_W^2}{R_A} \quad (13)$$

For the relative difference of the spectra compared to the channel voltage spectrum, $\approx 4kT_{eff}R_B$, we get the following estimations by Eq. (12 and 13):

Based on ref. [2] :

$$\frac{S^{[2]}_{ch,A-B}(f)}{S_{ch}(f)} \approx \frac{R_W}{R_B} \quad (14)$$

Based on Eq. (13) :

$$\frac{S_{ch,A-B}(f)}{S_{ch}(f)} \approx \frac{R_W^2}{R_A R_B} \quad (15)$$

For example, with $R_A = 100\,k\Omega$, $R_B = 1\,k\Omega$, $R_W = 100\,\Omega$, Eq. (14) yields 0.1 while Eq. (15) yields $10^{-4}$ which is 3 orders of magnitude lower. This means that, according to this analysis, Eve needs $10^3$ times longer measurement window to resolve the bit situation compared to the results of [2] because the uncertainty is mean-square estimation is inversely proportional with the measurement time window for band-limited white noise in the long-time limit.

## Conclusions

The research of secure communication is progressing by a continuous process of proposing new schemes, attempting to break/attacking them, followed by the development of defense mechanisms and/or modified schemes. We revisited the impact of finite resistance on the KLJN cipher and found that that the predicted information leak due to parasitic wire resistance is significantly less than it was previously estimated. In addition, the scheme can be improved to counteract such attacks because Alice and Bob can determine Eve's best guess based on the voltage difference at the two ends [6]. If these security measures are still insufficient, the same path can be followed as with quantum communicators: applying privacy amplification to the exchanged key.

## Acknowledgements

LBK is grateful to V. Makarov about information on quantum hacking.